\documentclass [aps,prb,twocolumn,showpacs] {revtex4}
\usepackage[final]{graphics}
\usepackage{amssymb}
\usepackage{amsfonts}
\usepackage{epsfig}

\begin{document}

\title{ Blume-Capel model on {\it directed} and {\it undirected} Small-World 
 Voronoi-Delaunay  random  lattices}

\author{F.P. Fernandes$^{1}$, F.W.S. Lima$^{1}$, and J. A. Plascak$^{2,3}$}
\affiliation{$^{1}$Departamento de F\'{\i}sica,
Universidade Federal do Piau\'{\i} , 64049-550, Teresina, PI, Brazil}
\affiliation{$^{2}$Departamento de F\'{\i}sica,
Universidade Federal de Minas Gerais, C. P. 702, 30123-970, Belo Horizonte, 
MG, Brazil}
\affiliation{$^3$ Center for Simulational Physics, 
University of Georgia, Athens, GA 30602, USA}

\begin{abstract}

The critical properties of the spin-$1$ two-dimensional Blume-Capel model
on {\it directed} and {\it undirected} random lattices with quenched connectivity
disorder is studied through Monte Carlo simulations. 
The critical temperature, as well as the critical point 
exponents are obtained. For the {\it undirected} case  this random 
system  belongs to the same universality
class as the regular two-dimensional  model. However, for the
{\it directed} random lattice one has a
second-order phase transition for $q<q_{c}$ and a first-order phase transition for $q>q_{c}$, 
where $q_{c}$ is the critical rewiring probability.
The critical exponents for $q<q_{c}$ was calculated and they do
not belong to the same universality class as the regular two-dimensional ferromagnetic model. 

\end{abstract}

\pacs {05.70.Ln, 05.50.+q, 75.40.Mg, 02.70.Lq}

\maketitle

\section{Introduction}

There has been recently an increasing interest in the study of networks
which are not usual  regular lattices. This growing interest, principally among physicists, 
comes mainly from the fact that they can model complex systems in real
world networks \cite{bara}. In special, small-world (SW) networks \cite{ws} have been
much studied as they furnish a simple way to attempt to describe complex
topologies such as social and economic organizations, the 
speed of disease spread, connectivities 
of computers, just to cite a few of them \cite{bara}. 

As is usual, a rigorous treatment to get the properties
of a real physical realization on such
complex systems is a quite difficult task. However, one expects that simple 
theoretical models, such as the discrete spin Ising model, could be able to capture the
main features of the actual complicated processes occurring on real
networks. In this direction, the spin-$1/2$ Ising model has been studied on SW 
networks and, for exchange interactions which are independent of the length
distance of the spins, one gets a second-order phase transition at a finite
critical temperature $T_c$ \cite{barrat,novo,hinc}. 

S\'anches et al. \cite{san} introduced the asymmetric directed SW networks and for an Ising
like spin system they showed that a first-order phase transition takes place when
the rewiring probability $q$ is greater than a critical value $q_c$. The same
result has been achieved by Lima et al. when treating the spin-$1,~ 3/2$ and $2$
Ising model on the same networks \cite{lima}.

In most of the above theoretical models the undirected, as well as the directed SW networks, 
have been constructed by taking, as an initial step, a regular pure hypercubic lattice
\cite{san}.
It will be very interesting, thus, to see what should be the behavior of SW networks built up
over an already random system, such as the Voronoi-Delaunay random lattices. 
So, in this paper,
we study the spin-$1$ Blume-Capel model on undirected and directed SW networks constructed
from an initial Voronoi-Delaunay random lattice. 
The purpose here is indeed two-fold: first, we would like
to see the effect of the disorder itself on the spin-$1$ model on this undirected lattice 
(since, for the spin-$3/2$ model \cite{wel} there has been a change in its universality 
class) and, secondly,
the behavior of spin models on SW networks obtained from random number of nearest neighbors. 
In the next section we present the SW networks, the
model and the simulation background. The results and
discussions are presented in the last section.

 \section{SW networks, Model and  Simulation}

{\it Undirected Voronoi-Delaunay random lattice} -  The Voronoi construction, 
or tessellation, for a given set of points
in the plane can be defined as follows \cite{christ}. Initially, for each point one
determines the polygonal cell consisting of the region of space
nearer to that point than any other point. Then one considers that the
two cells are neighboring when they possess an extremity in
common. From the Voronoi tessellation the dual
lattice can be obtained  by the following procedure:
$(i)$ when two cells are neighbors, a link is placed between the
two points located in the cells;
$(ii)$ from the links one obtains the triangulation of space
that is called the Delaunay lattice;
$(iii)$ the Delaunay lattice is dual to the Voronoi tessellation in
the sense that points corresponding to cells link to edges, and
triangles to the vertices of the Voronoi tessellation.

{\it Directed Voronoi-Delaunay SW network} - 
We use here the same process defined by S\'anchez et al. \cite{san} for directing 
the  Voronoi-Delaunay random lattice. In this case, we start with the  Voronoi 
construction, or tessellation, as described in the previous paragraph, by creating the
random two-dimensional lattice consisting of sites linked to their nearest neighbors
by both outgoing and incoming links. Then, with probability $q$, we reconnect
nearest-neighbor outgoing links to a different site chosen at random. After repeating
this process for every link, we are left with a network with a density $q$ of random 
two-dimensional lattice {\it directed} links. Therefore, with this procedure every site 
will have exactly the same amount of {\it undirected} random two-dimensional lattice
outgoing links   and a varying (random) number of incoming links. 

{\it Model} - We consider now the two-dimensional spin-$1$ Blume-Capel model on these
Poissonian random lattices. The  Blume-Capel Model is a generalization
of the standard Ising model \cite{kobe} and was originally proposed for spin-$1$ to 
account for first-order phase transition in magnetic systems \cite{BC,BC2}. The Hamiltonian
can be written as
\begin{equation}
H=-J\sum_{<i,j>}S_{i}S_{j}+\Delta\sum_{i}S_{i}^{2},
\end{equation}
where the first sum runs over all nearest-neighbor pairs of sites
(points in the Voronoi construction) and the spin-$1$ variables $S_{i}$ assume values 
$\pm 1,0$. In eq. (1) $J$ is the exchange coupling and $\Delta$ is the single ion
anisotropy parameter. The  second sum is 
taken over the $N$ spins on a $D$-dimensional lattice. 
The case where $S=1$ has been extensively studied by several approximate techniques
in two- and three-dimensions and its phase diagram is well established 
\cite{BC,BC2,BC3,BC4,BC5,BC6,BC7}. The case $S>1$ has also been investigated according
to several procedures \cite{BC8,BC9,BC10,BC11,BC12,BC13,landaupla}. 

{\it Simulations} - The simulations have been performed for $\Delta=0$, which is the 
simplest case and where we have some results in the literature to compare with. The  
lattices comprised different sizes with $N$ ranging from $N=250, 500, 1000, 2000,
4000, 8000$,  and $16000$, where $N$ is total umber
of sites. For simplicity,  the linear length of the system is defined here in terms
of the size of a regular two-dimensional lattice $L=N^{1/2}$.
For each system size quenched averages over the connectivity disorder are 
approximated by averaging over $R=100$ ($N=500$ to $4000$), $R=50$ ($N=8000$) and $R=25$ 
($N=16000$) independent realizations. For each
simulation we have started with a uniform configuration of spins
(the results are, however, independent of the initial configuration). 
We ran $2.52\times10^{6}$ Monte Carlo steps (MCS) per spin with $1.2\times10^{5}$ 
configurations  discarded for thermalization using the ``perfect" random-number 
generator \cite{nu}.
In both cases we have employed the heat bath algorithm and
for every $12$th MCS, the energy per spin,
$e=E/N$, and magnetization per spin, $m=\sum_{i}S_{i}/N$, were
measure and recorded in a time series file. From these thermodynamic quantities
the behavior of the transition can be analyzed.

From the series of the energy measurements we can compute, by re-weighting over a controllable
temperature interval $\Delta T$, the average energy, the specific heat, and the energy
fourth-order cumulant which are respectively given by
\begin{equation}
 u(K)=[<E>]_{av}/N,
\end{equation}
\begin{equation}
 C(K)=K^{2}N[<e^{2}>-<e>^{2}]_{av},
\end{equation}
\begin{equation}
 B(K)=[1-\frac{<e^{4}>}{3<e^{2}>^{2}}]_{av},
\end{equation}
where $K=J/k_BT$, with a normalized exchange $J=1$, and $k_B$ is the Boltzmann constant. 
In the above equations
$<...>$ stands for  thermodynamic averages and $[...]_{av}$
for  averages over the different realizations.
Similarly, we can derive from the magnetization measurements
the average magnetization, the susceptibility, and the magnetic
cumulants,
\begin{equation}
 m(K)=[<|m|>]_{av},
\end{equation}
\begin{equation}
 \chi(K)=KN[<m^{2}>-<|m|>^{2}]_{av},
\end{equation}
\begin{equation}
 U_{2}(K)=[1-\frac{<m^{2}>}{3<|m|>^{2}}]_{av},
\end{equation}
\begin{equation}
 U_{4}(K)=[1-\frac{<m^{4}>}{3<|m|>^{2}}]_{av}.
\end{equation}
Further useful quantities involving both the energy and magnetization are
their derivatives
\begin{equation}
 \frac{d[<|m|>]_{av}}{dK}=[<|m|E>-<|m|><E>]_{av},
\end{equation}
\begin{equation}
 \frac{d\ln[<|m|>]_{av}}{dK}=[\frac{<|m|E>}{<|m|>}-<E>]_{av},
\end{equation}
\begin{equation}
 \frac{d\ln[<|m^{2}|>]_{av}}{dK}=[\frac{<|m^{2}|E>}{<|m^{2}|>}-<E>]_{av}.
\end{equation}

In order to get the transition temperature, as well as to determine  the order of the 
transition of this model, we apply the finite-size scaling
(FSS) procedure. Initially, we search for the minima of the energy fourth-order cumulant
given by Eq.
(4). This quantity gives a qualitative, as well as a quantitative, description of the order
of the transition \cite{mdk}. 
For instance, it is known \cite{janke0} that this parameter takes a minimum
value $B_{min}$ at the effective transition temperature $T_{c}(N)$. One can 
also show \cite{kb}
that for a second-order transition, $\lim_{N\to \infty}$ $(2/3-B_{min})\rightarrow 0$, even at
$T_{c}$, while at a first-order transition the same limit measuring the same quantity is
small and $(2/3-B_{min})\neq 0$.

A more quantitative analysis can be carried out through the FSS of the 
energy fluctuation
$C_{max}$, the susceptibility maxima $\chi_{max}$, and the minima of the Binder cumulants 
$B$, $U_2$ and $U_4$. 

If the hypothesis of a first-order phase transition is correct, we should
then expect, for large systems sizes, an asymptotic FSS behavior of the form
\cite{wj,pbc},
\begin{equation}
C_{max}=a_{C} + b_{C}N +...
\end{equation}
\begin{equation}
\chi_{max}=a_{\chi} + b_{\chi}N +...
\end{equation}
\begin{equation}
B_{min}=a_{B_{i}} + b_{B_{i}}/N +...
%?? Aqui eu mudei  b_{B_{i}}N +...==>b_{B_{i}}/N +...
\end{equation}

On the other hand, if the hypothesis of a second-order phase transition is correct, we should
then expect, for large systems sizes, an asymptotic FSS behavior of the form
\begin{equation}
 C=C_{reg}+L^{-\alpha/\nu}f_{C}(x)[1+...],
\end{equation}
\begin{equation}
 [<|m|>]_{av}=L^{-\beta/\nu}f_{m}(x)[1+...],
\end{equation}
\begin{equation}
 \chi=L^{-\gamma/\nu}f_{\chi}(x)[1+...],
\end{equation}
\begin{equation}
 \frac{d \ln[<|m|^{p}>]_{av}}{dK}=L^{1/\nu}f_{p}(x)[1+...],
\end{equation}
%\begin{equation}
% \frac{dU_{2p}}{dK}=L^{1/\nu}f_{U_{2p}}(x)[1+...],
%\end{equation}
where $C_{reg}$ is a regular background term, 
 $\nu$, $\alpha$, $\beta$, and $\gamma$ are the usual critical
exponents, and $f_{i}(x)$ are FSS functions with
%\begin{equation}
$ x=(K-K_{c})L^{1/\nu}$
%\end{equation}
being the scaling variable, and the brackets $[1+...]$ indicate
corrections-to-scaling terms. 

In all cases, we estimated the error bars from the fluctuations 
among the different realizations. Note that these errors contain both, the average
thermodynamic error for a given realization and the theoretical
variance for infinitely accurate thermodynamic averages which are
caused by the variation of the quenched, random geometry of the lattices.

\section{Results and discussion}

\subsection{\bf undirected random two-dimensional lattice}

By applying standard re-weighting techniques to each of the $R$ time-series data
we first determined the temperature dependence of $C_{i}(K)$, $\chi_{i}(K)$,..., $i=1$,...,$R$,
in the neighborhood of the simulation point $K_{0}$.
Once the temperature dependence is known for each realization, we can easily compute
the disorder average, e.g., $C(K)=\sum^{R}_{i=1}C_{i}(K)/R$, and then determine the maxima 
of the averaged quantities, e.g., $C_{max}(K_{max})=max_{K}C(K)$. The variable $R$ 
represents the number of replicas in our simulations.

\begin{figure}[ht]
\includegraphics[clip,angle=-90,width=8.5cm]{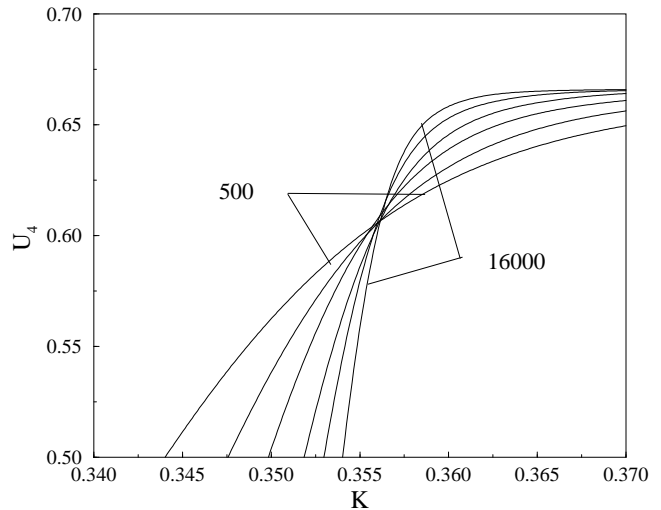}
\caption{\label{cum}
Fourth-order Binder cumulant as a function of $K$ for several values of the system 
size   $N=500,1000, 2000, 4000, 8000$ and $16000$.}
\end{figure}

In  order to estimate the critical temperature we calculate the second and fourth-order 
Binder cumulants given by  eqs. (7) and (8), respectively. It is well known that these
quantities are independent of the system size and should intercept at the critical 
temperature \cite{binder}. In Fig. \ref{cum} the fourth-order
Binder cumulant is shown as a function of the $K$ for several values of $N$. Taking the 
largest lattices we have  $K_{c}= 0.3560(5)$. To estimate $U^{*}_{4}$ we note that it
varies little at  $K_{c}$ 
so we have $U^{*}_{4}= 0.6058(6)$. From the second-order
cumulant (not shown)
we similarly get $K_{c}= 0.3561(4)$ and $U^{*}_{2}= 0.6416(3)$. One can see that 
the agreement of the critical temperature is quite good and  $U^{*}_{4}$ is definitely
close to the universal value $U^{*}_{4}\sim 0.61$  for the same model on the regular 
$2D$ lattice.
%
%
%%%%%%%%%%%%%%%%%%%%%%%% fig.1 %%%%%%%%%%%%%%%%%%%%%%%%%%%%%%%%%%%%%%%%%%%%%%%%
%%%%%%%%%%%%%%%%%%%%%%%%%%%%%%%%%%%%%%%%%%%%%%%%%%%%%%%%%%%%%%%%%%%%%%%%%%%%%%%%%
%
% 

\begin{figure}[ht]
\includegraphics[clip,angle=-90,width=8.5cm]{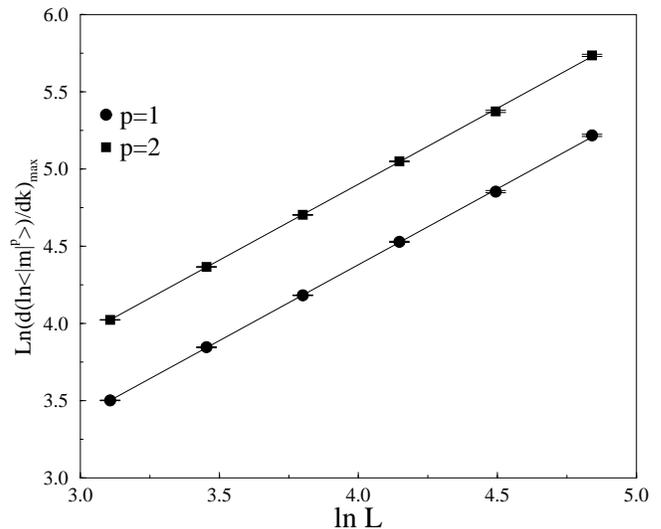}
\caption{\label{expnu}
Log-log plot of the maxima of the logarithmic derivative 
$\frac{d \ln[<|m|^{p}>]}{dK}$ versus  the lattice size $L=N^{1/2}$ for 
$p=1$ (circle) and $p=2$ (square). The solid lines are the best linear fits. 
}
\end{figure}

The correlation length exponent can be estimated from the derivatives given by
eq. (18). Figure \ref{expnu} shows the maxima of the logarithm derivatives as a 
function of the logarithm of the lattice size $L$ for $p=1$ and $p=2$. From the linear
fitting one gets $\nu=0.984(6)$ ($p=1$) and  $\nu=0.983(5)$ ( $p=2$), which is again
next to  the regular lattice exponent $\nu=1$.
%
%
%%%%%%%%%%%%%%%%%%%%%%%% fig.2 %%%%%%%%%%%%%%%%%%%%%%%%%%%%%%%%%%%%%%%%%%%%%%%%
%%%%%%%%%%%%%%%%%%%%%%%%%%%%%%%%%%%%%%%%%%%%%%%%%%%%%%%%%%%%%%%%%%%%%%%%%%%%%%%%%
%

\begin{figure}[ht]
\includegraphics[clip,angle=-90,width=8.5cm]{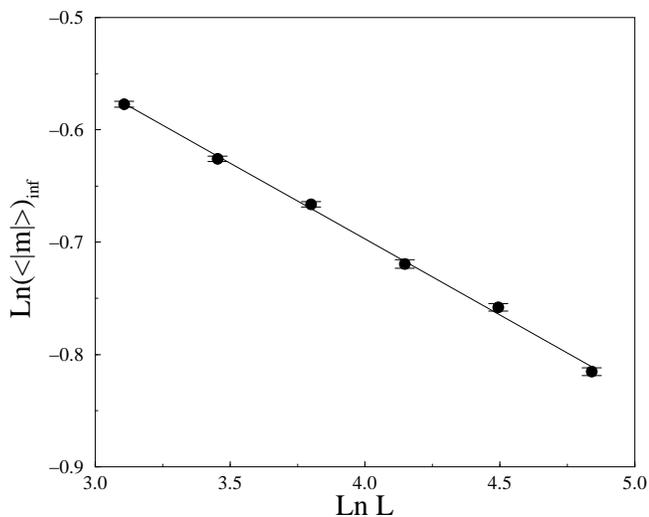}
\caption{\label{mag}
Plot of the logarithm of the modulus of the magnetization at the
inflection point as a function of the logarithm of  $L=N^{1/2}$. The solid line is 
the best linear fit.
}
\end{figure}

\begin{figure}[ht]
\includegraphics[clip,angle=-90,width=7.9cm]{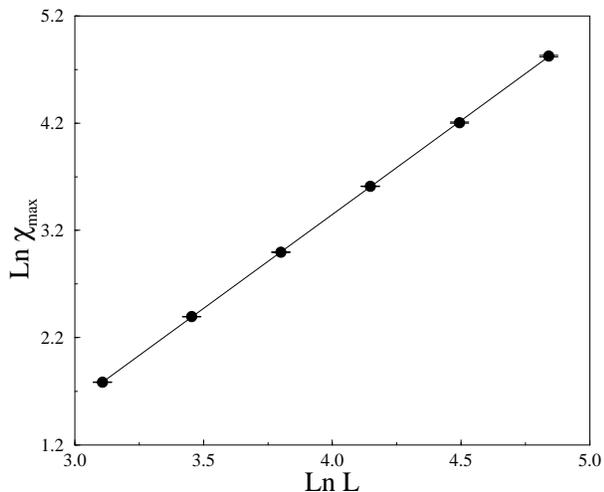}
\caption{\label{sus}
 Log-log plot of the susceptibility maxima $\chi_{max}$ as a function of
the logarithm of  $L=N^{1/2}$. The solid line is the best linear fit.}
\end{figure}

In order to go further in our analysis we also computed the modulus of the magnetization
at the inflection point and the maximum of the magnetic susceptibility. The logarithm of
these quantities as a function of the logarithm of  $L$ are presented in Figures  
\ref{mag} and \ref{sus},
respectively. A linear fit of these data gives  $\beta/\nu=0.135(9)$ from the magnetization
and  $\gamma/\nu=1.751(4)$ from the susceptibility which should be compared to
 $\beta/\nu=0.125$ and  $\gamma/\nu=1.75$ obtained for a regular $2D$ lattice. One can see
that for this undirected random lattice we get the same universal critical behavior as the 
regular lattice model, in the same way as has been reported for the diluted spin-$1/2$
Ising model \cite{paulo}.

%
%
%%%%%%%%%%%%%%%%%%%%%%%% fig.3 %%%%%%%%%%%%%%%%%%%%%%%%%%%%%%%%%%%%%%%%%%%%%%%%
%%%%%%%%%%%%%%%%%%%%%%%%%%%%%%%%%%%%%%%%%%%%%%%%%%%%%%%%%%%%%%%%%%%%%%%%%%%%%%%%%
%
%
%
%%%%%%%%%%%%%%%%%%%%%%%% fig.4 %%%%%%%%%%%%%%%%%%%%%%%%%%%%%%%%%%%%%%%%%%%%%%%%
%%%%%%%%%%%%%%%%%%%%%%%%%%%%%%%%%%%%%%%%%%%%%%%%%%%%%%%%%%%%%%%%%%%%%%%%%%%%%%%%%
%

%%%%%%%%%%%%%%%%%%%%%%%%%%%%%%%%%%%%%%%%%%%%%%%%%%%%%%%%%%%%%%%%%%%%%%%%%%%%%%%%%

\subsection{ directed random two-dimensional lattice}

\begin{figure}[ht]
\includegraphics[clip,angle=0,width=8.0cm]{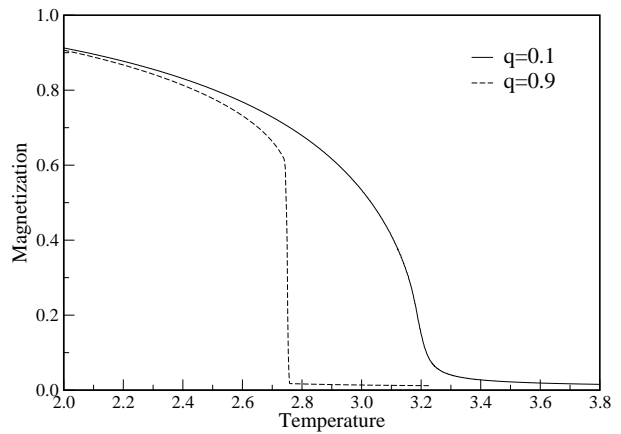}
\caption{\label{mag1}
Magnetization  as a function of the temperature, for
$N=16000$ sites. It is clear a second-order phase transition for $q=0.1$, and a first-order 
phase   transition for $q=0.9$, respectively.}
\end{figure}

\begin{figure}[ht]
\includegraphics[clip,angle=0,width=8.0cm]{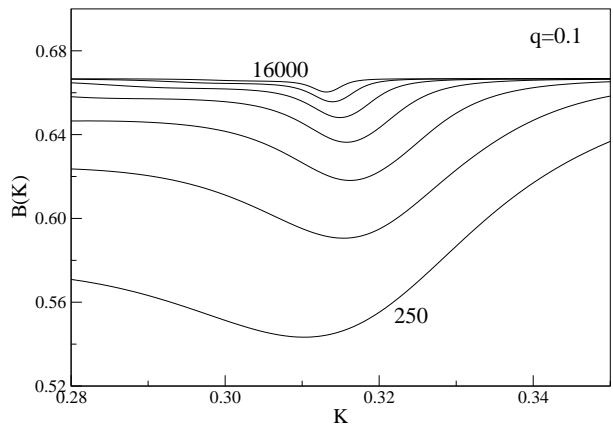}
\caption{\label{be1}
Energetic Binder cumulant as a function of the reduced temperature $K$ for $q=0.1$ and 
lattice sizes $N=250,500,1000,2000,4000,8000$, and $16000$ from bottom to top.
}
\end{figure}

\begin{figure}[ht]
\includegraphics[clip,angle=0,width=8.0cm]{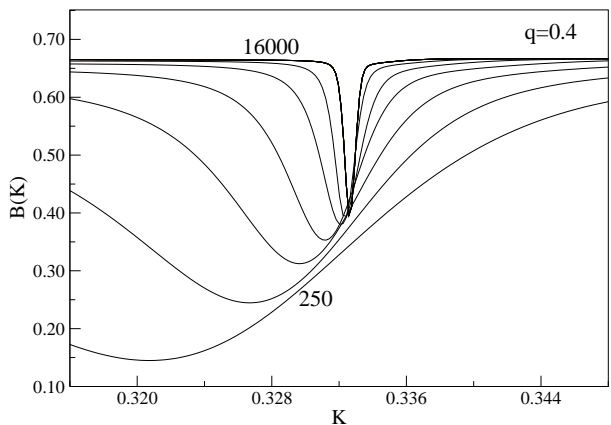}
\caption{\label{be2}
The same display the Fig. \ref{be1}, but now for $q=0.4$ and
% ?? Here eu mudei $p=0.4$ ==> $q=0.4$
size $N=250,500,1000,2000,4000,8000$, and $16000$ from bottom to top.
}
\end{figure}

In Fig. \ref{mag1} we show the dependence of the magnetization as a function  of the 
temperature, obtained
from simulations on directed random two-dimensional lattice,  with $N=16000$ sites, and two
values of probability $q$, namely $q=0.1$ for second-order phase transition and $q=0.9$ for
first-order phase transition, respectively.
%%%%%%%%%%%%%%%%%%%%%%%% fig.5 %%%%%%%%%%%%%%%%%%%%%%%%%%%%%%%%%%%%%%%%%%%%%%%%
%%%%%%%%%%%%%%%%%%%%%%%%%%%%%%%%%%%%%%%%%%%%%%%%%%%%%%%%%%%%%%%%%%%%%%%%%%%%%%%%%
%
The energetic Binder cumulant as a function of the reduced temperature $K$  for $q=0.1$, 
and different lattice
sizes, is  displayed in the Figure \ref{be1}, where a characteristic second-order phase 
transition is observed. 
%
%%%%%%%%%%%%%%%%%%%%%%%% fig.6a %%%%%%%%%%%%%%%%%%%%%%%%%%%%%%%%%%%%%%%%%%%%%%%%
%%%%%%%%%%%%%%%%%%%%%%%%%%%%%%%%%%%%%%%%%%%%%%%%%%%%%%%%%%%%%%%%%%%%%%%%%%%%%%%%%
%
On the other hand, in the Figure \ref{be2} we display the same plot as in Figure \ref{be1}, 
but now for $q=0.4$,
where we can see that a first-order transition is present for this value 
of the rewiring probability.
%
%%%%%%%%%%%%%%%%%%%%%%%% fig.6b %%%%%%%%%%%%%%%%%%%%%%%%%%%%%%%%%%%%%%%%%%%%%%%%
%%%%%%%%%%%%%%%%%%%%%%%% fig.7 %%%%%%%%%%%%%%%%%%%%%%%%%%%%%%%%%%%%%%%%%%%%%%%%

\begin{figure}[ht]
\includegraphics[clip,angle=0,width=8.0cm]{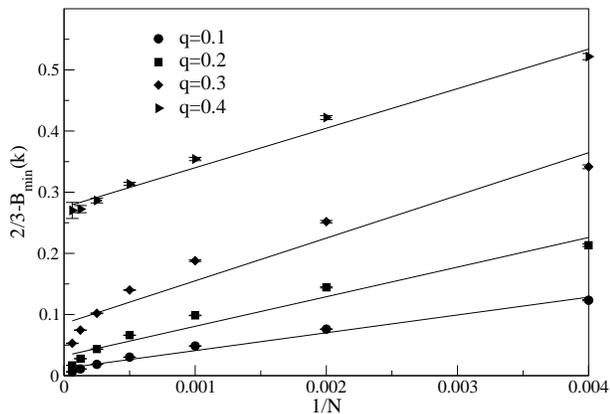}
\caption{\label{bim}
Plot of $2/3-B_{min}$ at $T_{c}$ as a function of logarithm of $1/N$ for several values of the system size  $N=250,500,1000, 2000, 4000, 8000$ and $16000$, and $q=0.1, 0.2$, $0.3$, and $0.4$.}
\end{figure}

In Fig.\ref{bim}, the difference $2/3-B_{min}$ is shown as a function of the
parameter $1/N$ for different probabilities $q$. For values  $q<q_{c}\approx 0.35$, 
a second-order transition takes place since in the $\lim_{N\to \infty}$ $(2/3-B_{i,min})=0$, 
even at $T_{c}$. However, for $q>q_c$ a first-order transition is observed, because one has
$(2/3-B_{i,min})\neq0$. For instance, for $q=0.4$ we get $(2/3-B_{min})=0.273(3)$. 
% ?? Here eu mudei $p=0.4$ ==> $q=0.4$ 
% ?? Here eu mudei $(2/3-B_{i,min})=0.273(3)$ ==>$(2/3-B_{min})=0.273(3)$ eu removi
% ??porque estamos falando da figura que é agora uma média dos B_{i,min} e agora está de
% ??acordo com a figura

\begin{figure}[ht]
\includegraphics[clip,angle=0,width=8.0cm]{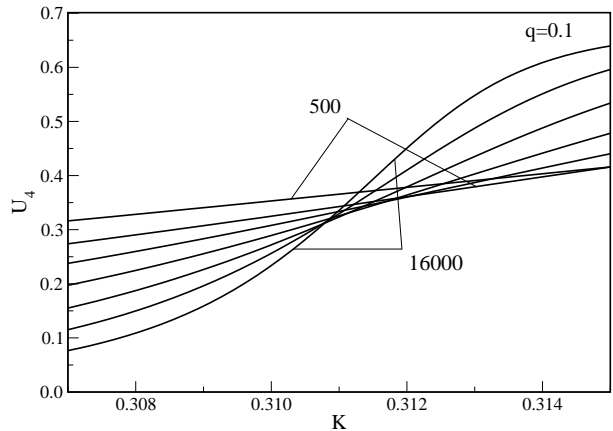}
\caption{\label{cum1}
Fourth-order Binder cumulant as a function of $K$ for several values of the system 
size   $N=500,1000, 2000, 4000, 8000$ and $16000$, and $q=0.1$.}
\end{figure}

In Fig. \ref{cum1} the fourth-order
Binder cumulant is shown as a function of the $K$ for several values of $N$ and $q=0.1$. 
Taking the 
largest lattices we have  $K_{c}= 0.3109(6)$. To estimate $U^{*}_{4}$ we note that it
varies little at  $K_{c}$ 
so we have $U^{*}_{4}= 0.326(4)$. From the second-order
cumulant (not shown) we similarly get $K_{c}= 0.31078(3)$ and $U^{*}_{2}=0.538(2) $.
For other values of $q$ we have: for $q=0.2$, $K_{c}= 0.3114(6)$ and $U^{*}_{4}= 0.326(5)$
and  $K_{c}= 0.31164(5)$ and $U^{*}_{2}=0.534(7)$; for $q=0.3$, $K_{c}= 0.32154(8)$ and 
$U^{*}_{4}= 0.327(5)$, and $K_{c}= 0.32137(5) $ and $U^{*}_{2}= 0.556(4)$. One can see that 
the agreement of the critical temperature is quite good and  $U^{*}_{4}$ is definitely
different from the universal value $U^{*}_{4}\sim 0.61$  for the same model on the regular 
$2D$ lattice and  undirected Voronoi-Delaunay random lattice.

%%%%%%%%%%%%%%%%%%%%%%%% fig.8 %%%%%%%%%%%%%%%%%%%%%%%%%%%%%%%%%%%%%%%%%%%%%%%%
%%%%%%%%%%%%%%%%%%%%%%%%%%%%%%%%%%%%%%%%%%%%%%%%%%%%%%%%%%%%%%%%%%%%%%%%%%%%%%%%%
%
%

\begin{figure}[ht]
\includegraphics[clip,angle=0,width=8.0cm]{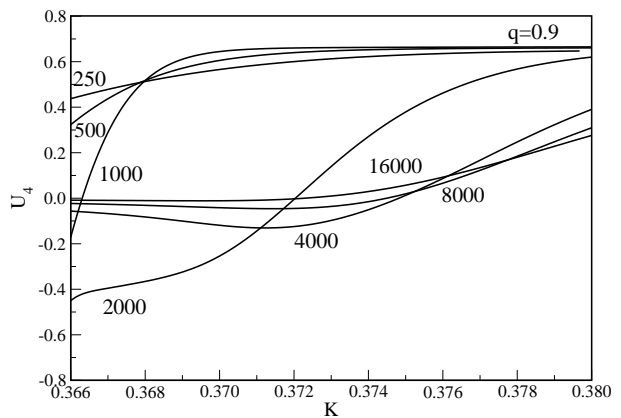}
\caption{\label{cum2}
Fourth-order Binder cumulant as a function of $K$ for several values of the system 
size   $N=250,500,1000, 2000, 4000, 8000$ and $16000$, and $q=0.9$.}
\end{figure}

Fig. \ref{cum2} displays the behavior of the magnetization fourth-order cumulant for $q=0.9$
in a narrow range of $K$, 
where one can note a different behavior leading to a first-order phase transition.
%%%%%%%%%%%%%%%%%%%%%%%% fig.8 %%%%%%%%%%%%%%%%%%%%%%%%%%%%%%%%%%%%%%%%%%%%%%%%
%%%%%%%%%%%%%%%%%%%%%%%%%%%%%%%%%%%%%%%%%%%%%%%%%%%%%%%%%%%%%%%%%%%%%%%%%%%%%%%%%
%

\begin{figure}[ht]
\includegraphics[clip,angle=0,width=8.0cm]{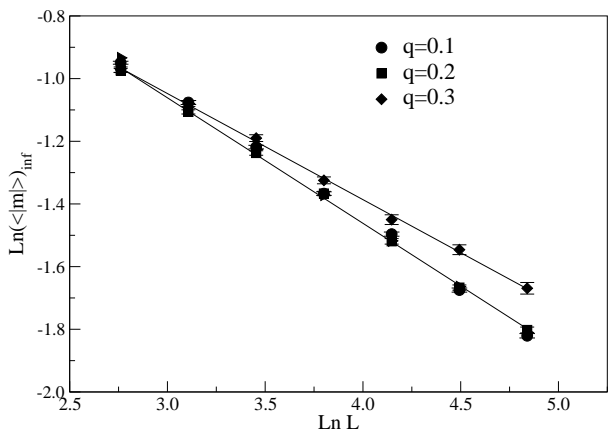}
\caption{\label{bim1}
Plot of the logarithm of the modulus of the magnetization at the
inflection point as a function of  $L=N^{1/2}$ for several values of the system size   
$N=250,500,1000, 2000, 4000, 8000$ and $16000$, and $q=0.1, 0.2,$  and $0.3$.}
\end{figure}

In our analysis we also computed the modulus of the magnetization
at the inflection point, $T=T_{c}$ for $q=0.1, 0.2,$  and $0.3$ . The logarithm of 
these quantities as a function of the logarithm of  $L$ are presented in Figure  
\ref{bim1}. A linear fit of these data gives  $\beta/\nu=0.421(7), 0.400(5)$ and
$0.338(5)$ from the magnetization for $q=0.1,0.2$, and $0.3$, respectively. 
These results are totally different of  $\beta/\nu=0.125$ obtained for a regular $2D$ lattice.

%%%%%%%%%%%%%%%%%%%%%%%% fig.9 %%%%%%%%%%%%%%%%%%%%%%%%%%%%%%%%%%%%%%%%%%%%%%%%
%%%%%%%%%%%%%%%%%%%%%%%%%%%%%%%%%%%%%%%%%%%%%%%%%%%%%%%%%%%%%%%%%%%%%%%%%%%%%%%%%
%

\begin{figure}[ht]
\includegraphics[clip,angle=0,width=8.0cm]{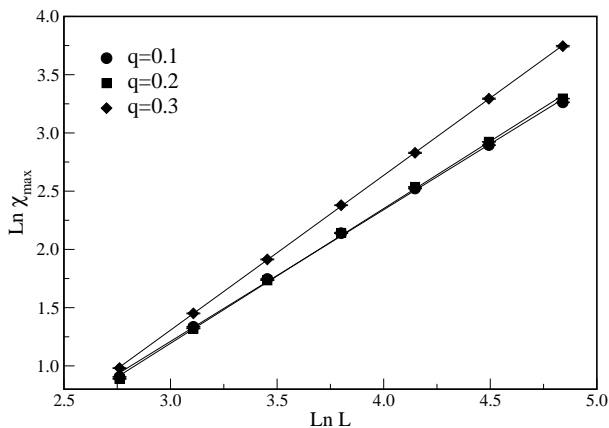}
\caption{\label{svt}
Logarithmic of the susceptibility at $T_{c}$ as a function of the logarithmic of 
$L=N^{1/2}$ for several values of the system 
size   $N=250,500,1000, 2000, 4000, 8000$ and $16000$, and $q=0.1, 0.2,$  and $0.3$.}
\end{figure}

In Fig. \ref{svt} the logarithm plot of the susceptibility at $T_{c}$ as a 
function of the logarithm of $L=N^{1/2}$ is presented. A linear fit of these data gives, 
at $T=T_{c}$, $\gamma/\nu=1.101(4), 1.138(5)$ and $1.312(7)$. Similarly, from the maximum of the 
magnetic susceptibility (not shown) one gets  $\gamma/\nu=1.130(14), 1.156(12)$ and $1.312(3)$ 
for $q=0.1, 0.2$ and $0.3$, respectively, which again is totally 
different of $\gamma/\nu=1 .75$ obtained for a regular $2D$ lattice.

%
%%%%%%%%%%%%%%%%%%%%%%%% fig.10 %%%%%%%%%%%%%%%%%%%%%%%%%%%%%%%%%%%%%%%%%%%%%%%%
%%%%%%%%%%%%%%%%%%%%%%%%%%%%%%%%%%%%%%%%%%%%%%%%%%%%%%%%%%%%%%%%%%%%%%%%%%%%%%%%%
%

\begin{figure}[ht]
\includegraphics[clip,angle=0,width=8.0cm]{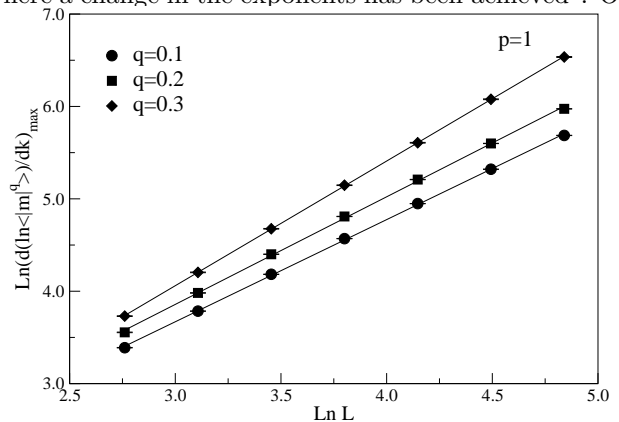}
\caption{\label{expnu1}
Log-log plot of the maxima of the logarithmic derivative 
$\frac{d \ln[<|m|^{p}>]}{dK}$ versus  the lattice size $L=N^{1/2}$ for 
$p=1$ and several values of the system 
size   $N=250,500,1000, 2000, 4000, 8000$ and $16000$, and $q=0.1, 0.2,$  and $0.3$.}
\end{figure}

Finally, in order to calculate the exponents $1/\nu$ ,  we use the maxima of the 
logarithmic derivative 
as defined in Eq. (18). A plot of this quantity versus  the lattice size $L=N^{1/2}$ for 
$p=1$ is shown in Figure \ref{expnu1}. From this figure one gets  
$1/\nu =1.105(8)$, $1.164(11)$  and $1.349(3)$. Similarly, for $p=2$ (not shown) we have 
$1/\nu =1.107(7), 1.167(11)$ and $1.349(3)$ for $q=0.1, 0.2$ and $0.3$, respectively, which 
are also again  different from $1/\nu =1 $ obtained for a regular $2D$ lattice.
%%%%%%%%%%%%%%%%%%%%%%%% fig.11 %%%%%%%%%%%%%%%%%%%%%%%%%%%%%%%%%%%%%%%%%%%%%%%%
%%%%%%%%%%%%%%%%%%%%%%%%%%%%%%%%%%%%%%%%%%%%%%%%%%%%%%%%%%%%%%%%%%%%%%%%%%%%%%%%%

Thus, from the above results, there is a strong indication that the spin-1 Blume-Capel
model on a {\it undirected} Voronoi lattice is in same universality class as its regular 
lattice counterpart, in contrast with the spin-$3/2$ model where a change in the exponents
has been achieved \cite{wel}. On the other hand,
on a {\it directed} SW Voronoi lattice, the exponents here obtained for $q<q_{c}$, where a
second order phase transition is obtained in two dimensions, are different from the 
spin-$1$ Blume-Capel model on regular $2D$ lattices. Therefore, for $q<q_{c}$ 
not only the exponents, but also the fourth-order cumulants, show a strong indication 
that this model is not in the same universality class than its regular $2D$ lattice,
with varying exponents as a function of $q$. 
For $q>q_{c}$, we have a first-order phase transition.

We believe that the above behavior should be 
qualitatively the same for crystal field $\Delta<\Delta_t$, where $\Delta_t$ is
the value where one has a tricritical point in the model on a regular lattice. However,
there are still some open questions, for instance:
i) the effect of the nearest-neighbor disorder on the undirected lattice for 
$\Delta>\Delta_t$, i.e. 
in the range of a first-order transition. One can ask whether that will result 
on a second-order transition with strong
violation of universality, as happens in the regular lattice random-bond model studied
by Malakis et al. \cite{malakis}; ii) the effect of large crystal field on the SW  networks
where a first-order already takes place on the regular lattice. Work in this direction is
now in progress.
 
%Then, all our results agree with the Harris-Luck
%criterion for  Voronoi-Delaunay random lattices.

\end{document}